\documentclass{article}
\usepackage{kr}

\usepackage{times}
\usepackage{soul}
\usepackage{url}
\usepackage[hidelinks]{hyperref}
\usepackage[utf8]{inputenc}
\usepackage[small]{caption}
\usepackage{graphicx}
\usepackage{amsmath}
\usepackage{amsthm}
\usepackage{booktabs}
\usepackage{algorithm}
\newtheorem{example}{Example}
\newtheorem{theorem}{Theorem}

\usepackage{algpseudocode}
\usepackage{listings}
\newtheorem{definition}{Definition}

\usepackage[all]{nowidow}

\title{Dispute Resolution in Peer Review with Abstract Argumentation and OWL DL}

\author{%
 Ildar Baimuratov$^1$\and
 Elena Lisanyuk$^{2,3}$\and
 Dmitry Prokudin$^{4}$\and
\affiliations
$^1$L3S Research Center, Leibniz University Hannover, Germany\\
$^2$Institute of Philosophy, Russian Academy of Sciences, Russia\\
$^3$National Research University Higher School of Economics, Russia\\
$^4$St Petersburg State University, Russia\\
\emails
ildar.baimuratov@l3s.de,
elisanyuk@hse.ru,
d.prokudin@spbu.ru
}

\begin{document}

\maketitle

\begin{abstract}
The peer review process for scientific publications faces significant challenges due to the increasing volume of submissions and inherent reviewer biases. While artificial intelligence offers the potential to facilitate the process, it also risks perpetuating biases present in training data. This research addresses these challenges by applying formal methods from argumentation theory to support transparent and unbiased dispute resolution in peer review. Specifically, we conceptualize scientific peer review as a single mixed argumentative dispute between manuscript authors and reviewers and formalize it using abstract argumentation frameworks. We analyze the resulting peer review argumentation frameworks from semantic, graph-theoretic, and computational perspectives, showing that they are well-founded and decidable in linear time. These frameworks are then implemented using OWL DL and resolved with reasoning engines. We validate our approach by annotating a corpus of scientific peer reviews with abstract argumentation frameworks and applying a proof of concept to resolve the annotated disputes. The results demonstrate that integrating our method could enhance the quality of published work by providing a more rigorous and systematic approach to accounting reviewer arguments.
\end{abstract}
\section{Introduction}


The review process for scientific publications is becoming increasingly complex due to the rapid growth in submission volumes. In 2013-2019, the annual increase in the number of manuscripts submitted to peer-reviewed journals has been an unprecedented 6.1\%, with a significant increase in the number of rejections \cite{Fire2019}. More than 15 million hours are spent each year reviewing manuscripts that are initially rejected and subsequently resubmitted to other journals \cite{Huisman2017}. The accompanying reinforcement of pre-existing biases in academia, such as ``first impression'' bias, the Dr. Fox effect, ideological and theoretical biases, as well as language and social identity bias further complicate the peer review process \cite{Lee2013}. Other challenges,  faced by the peer review process range from the selfish or competitive rejection of high-quality papers to the acceptance of low-quality manuscripts without careful validation \cite{Andrea2017}. Several initiatives are exploring the use of artificial intelligence (AI) to tackle challenges in scientific peer review \cite{kousha2024artificial}. However, AI deployment brings concerns about reliability and the potential reinforcement of existing biases \cite{Checco2021}.




Our study addresses these issues by focusing on formal methods for representing and analyzing argumentation in peer review, emphasizing explainability and unbiasedness in the evaluation process. In particular, we conceptualize peer review as a single mixed argumentative dispute \cite{Eemeren2004} between the authors of a manuscript submitted for publication and the reviewers evaluating it. We formalize this process using abstract argumentation frameworks \cite{Dung1995} and analyze semantic, graph-theoretic, and computational properties of the resulting frameworks. To validate our approach, we annotate a corpus of scientific peer reviews with abstract argumentation frameworks, then generate corresponding OWL representations and resolve them using OWL reasoning, following the method described in \cite{Baimuratov2023}. Finally, we compute and analyze statistics of the resolved peer review argumentation frameworks.

Thus, the contributions of this paper are: 1) modeling argumentative disputes in peer review with abstract argumentation frameworks and OWL, 2) analyzing peer review argumentation frameworks with respect to their semantic, graph-theoretic and computational properties, and 3) evaluating the approach on a corpus of peer reviews annotated with abstract argumentation frameworks.

The remainder of the paper is organized as follows: \autoref{sec:rel} reviews related work, \autoref{sec:back} provides definitions for key concepts and outlines the method for representing abstract argumentation frameworks in OWL, \autoref{sec:method} argues for the formalization of peer review using abstract argumentation frameworks, analyzes their theoretical properties, and demonstrates their representation in OWL DL. Finally, \autoref{sec:eval} presents the evaluation of our method before the conclusion in \autoref{sec:conc}.
\section{Related Work}
\label{sec:rel}

This section reviews the existing literature on automating dispute resolution in peer review through abstract argumentation frameworks. Specifically, it covers AI applications in peer review, argument representation methods and abstract argumentation solvers.

\subsection{Peer Review and Artificial Intelligence}

Experimental initiatives aimed at significantly transforming the peer review process are currently under development. One prominent approach is open peer review, which offers an alternative model of interaction between authors and reviewers \cite{Ross-Hellauer2017}. A key example is the OpenReview project \cite{Soergel2013}, which seeks to enhance transparency in scholarly communication. A variety of alternative systems and online review platforms have been explored in \cite{Tennant2017}, ranging from voting mechanisms similar to those on Reddit to innovative blockchain-based models.

AI shows considerable promise in addressing challenges within the peer review process. The study \cite{Price2017} demonstrates that AI and machine learning can effectively automate and enhance several stages of the review process, including the assignment of articles (or grant applications) to appropriate reviewers. \cite{Ghosal2019} explore the impact of reviewer sentiment embedded in review texts as a predictor of review outcomes. The PEERAssist system \cite{Bharti2021} leverages a cross-attention mechanism between the full article text and the review text to predict reviewer decisions. \cite{Mrowinski2017} show that evolutionary algorithms can significantly optimize editorial strategies, accelerating the review process and reducing the burden on editors. Other notable examples include Statcheck \cite{Nuijten2020}, software that verifies the consistency of authors' statistics with a focus on $p$-values; Penelope.ai\footnote{https://www.penelope.ai/}, a commercial platform that ensures citations and manuscript structures meet journal guidelines; and StatReviewer\footnote{http://blogs.biomedcentral.com/bmcblog/2016/05/23/peerless-review-automating-methodological-statistical-review}, which checks the validity of statistics and methods in manuscripts.

The ethical challenges posed by such approaches are often related to the risks of reproducing biases within AI systems. \cite{Checco2021} discuss the potential and limitations of employing AI to support human decision-making in the quality assurance and peer review of scientific research. Their findings suggest that AI can reinforce existing biases in the peer review process. \cite{kousha2024artificial} conclude that AI shows promise in areas such as reviewer selection and initial quality control of submitted manuscripts, while its effectiveness in supporting the peer review process has yet to be clearly demonstrated.

\subsection{Argument Representation}

One of the foundational frameworks in argumentation theory are Dung's abstract argumentation frameworks (AAF) \cite{Dung1995}. Computational models designed to solve such frameworks are evaluated through the International Competition on Computational Models of Argumentation\footnote{https://argumentationcompetition.org/index.html} (ICCMA). In addition to standard argumentation frameworks, ICCMA 2019 introduced a dynamic track \cite{bistarelli2018containerisation}, which involves processing a sequence of changes to the attack structure of an AAF. ICCMA 2021 further expanded the competition with two new tracks: 1) on assumption-based argumentation \cite{bondarenko1997abstract} and 2) on approximate algorithms. While abstract argumentation solvers are an indispensable component of theories of argumentation, they do not address the nature of individual arguments or guide the modeling of real-world argumentation problems. For example, consider the argument representation used in ICCMA, as illustrated in \autoref{ex:iccma}. This format lacks mechanisms for representing the content or provenance of the arguments, limiting its ability to capture the full context of argumentative interactions.

\begin{example}
\label{ex:iccma}
An AAF with a set of arguments $A=\{a, b, c, d, e\}$ and a set of attacks $R=\{(a, b), (b, d), (d, e), (e, d), (e, e)\}$, assuming the indexing $a=1$, $b=2$, $c=3$, $d=4$, $e=5$, in ICCMA is represented as follows:\\
\texttt{p af 5\\
1 2\\
2 4\\
4 5\\
5 4\\
5 5}
\end{example}



To address these limitations, more sophisticated argument representations have been proposed. One widely used format is the Argument Interchange Format (AIF) \cite{Chesnevar2006}, which is based on the concept of argumentation schemes from \cite{Walton2008} and is designed to facilitate data exchange between different argumentation tools and applications. Various implementations of AIF exist, for example, \cite{Rahwan2009} proposed an AIF implementation using OWL. The online database AIFdb \cite{Lawrence2012} was created to store annotated argumentative texts. The AIF format is also used in the OVA tool \cite{Reed2014} for analyzing and annotating natural language argumentation.
Alternatively, ASPIC+ \cite{modgil2014aspic+} is a structured argumentation framework that enables modeling conflicts between arguments and assumes three ways of attacking: 1) by challenging their uncertain premises, 2) by attacking their defeasible inferences, and 3) by disputing the conclusions drawn from defeasible inferences. Another notable format is Argdown\footnote{https://argdown.org/}, an argument markup language inspired by Markdown, implemented using a context-free grammar and parser.

Currently, frameworks such as AIF, ASPIC+, and Argdown support the argumentation representation, the generation of argument maps and the classification of arguments using argumentation schemes. However, they lack tools for computational dispute resolution. Moreover, ASPIC+ and Argdown do not offer implementations based on Semantic Web standards, such as RDF and OWL, which would significantly enhance the interoperability and machine interpretability of argument representations. This highlights a notable gap between advanced argument representation methods and argumentation solvers. Only a few studies have sought to bridge this gap. For example, \cite{moguillansky2016generalized} explored the encoding of abstract argumentation within ALC description logic to enable reasoning over inconsistent ontologies, though they did not provide an implementation. In contrast, \cite{Baimuratov2023} proposed a promising method for representing AAF in OWL DL, enabling dispute resolution via OWL reasoning while preserving the advantages of OWL-based knowledge modeling. The present research builds on this approach to model disputes within the context of peer review.

\section{Background}
\label{sec:back}

In this section, we provide the background required to define our method for facilitating dispute resolution in peer review, including the abstract argumentation frameworks we use to formalize disputes in peer review, and the method for representing abstract argumentation frameworks in OWL DL we adopt from \cite{Baimuratov2023}.

\subsection{Abstract Argumentation Frameworks}

In this study, we use Dung's abstract argumentation frameworks \cite{Dung1995} to model disputes in peer review. These frameworks take an unstructured argument as the atomic unit of an argumentative dispute, and a pair consisting of an argument and its objection as the smallest molecular unit. Argumentation logic, which employs abstract argumentation frameworks, treats the process of persuasion as an exchange of reasoning between rational agents, and ``studies the criteria ... that determine the extent to which it is reasonable to accept logically deduced conclusions, even though some of these conclusions are established by non-deductive reasoning'' \cite{Prakken2002}. To model varying degrees of such reasonableness, argumentation logics rely on computational semantics. Widely used versions of these semantics include extension semantics, initially proposed for certain non-monotonic logics \cite{Kakas1992}, as well as preference semantics \cite{Amgoud2018,Grossi2019}, and labeling semantics \cite{Pollock1995,Caminada2009}.

In abstract argumentation frameworks, the process of argument confirmation is represented through a graph where arguments are connected by a binary asymmetric attack relation. This relation symbolizes criticism or counterargumentation, serving as the fundamental mechanism for interactions between arguments in a dispute. Thus, any argumentation framework can be represented as a directed graph.

\begin{definition}
An \textbf{argumentation framework} $AF$ is a pair
\begin{equation*}
    AF = <A, R>,
\end{equation*}
where $A$ is a set of arguments and $R\subseteq A\times A$ is the attack relation.
\end{definition}

We say that an argument $\alpha\in A$ attacks an argument $\beta\in A$, or that $\beta$ is attacked by an argument $\alpha$ if $(\alpha, \beta)\in R$. Additionally, we say that a set of arguments $S\subseteq A$ attacks $\alpha$, or that $\alpha$ is attacked by $S$ if some argument $\beta\in S$ attacks $\alpha$:
\begin{equation*}
    attacks(S, \alpha)\equiv_{def}\exists \beta\in S\ (\beta, \alpha)\in R.
\end{equation*}

In order to define outcomes and then solutions to disputes, we first utilize the notion of a conflict-free set of arguments.
\begin{definition}
A set of arguments $S$ is called \textbf{conflict-free} if there are no arguments $\alpha$ and $\beta$ in $S$ such that $(\alpha, \beta)\in R$.
\begin{equation*}
    cf(S)\equiv_{def}\forall\beta\in S\ \neg attacks(S, \beta).
\end{equation*}
\end{definition}

However, a conflict-free set of arguments alone is not sufficient to define a solution to a single mixed dispute. The smallest degree of reasonableness that characterizes the arguments that can persuade a rational agent is an \textit{admissible} subset, consisting of \textit{acceptable} arguments.

\begin{definition}
An argument $\alpha\in A$ is acceptable on a set of arguments $S$ only if, whenever it is attacked by an argument $\beta$, $S$ attacks $\beta$.
\begin{multline*}
    acc(\alpha, S)\equiv_{def}\forall\beta\in A\ (\beta, \alpha)\in R \implies\\
    \implies attacks(S, \beta).
\end{multline*}
\end{definition}

Now we can define an admissible set of arguments.
\begin{definition}
A conflict-free set of arguments $S$ is \textbf{admissible} only if every argument in $S$ is acceptable with respect to $S$.
\begin{equation*}
    adm(S)\equiv_{def}cf(S)\wedge\forall\alpha\in S\ acc(\alpha, S)
\end{equation*}
\end{definition}

To determine whether an argument can be accepted, either individually or together with others, various acceptance semantics are introduced. These semantics allow for the computation of sets of arguments, known as extensions, including preferred, stable, complete and grounded extensions.

\begin{definition}
A set $S$ is a \textbf{preferred extension} of an argumentation framework $AF$ if it is a maximal (with respect to the set-theoretic inclusion) admissible set in $AF$.
\begin{equation*}
    pref(S)\equiv_{def}adm(S)\wedge(\forall T\subseteq A\ S\subset T\implies\neg adm(T)).
\end{equation*}
\end{definition}

\begin{definition}
A set of arguments $S$ is a \textbf{stable extension} only if $S$ is conflict-free and attacks each argument that does not belong to $S$.
\begin{multline*}
    stable(S)\equiv_{def}cf(S)\wedge(\forall\alpha\in A\ (\alpha\not\in S)\implies\\
    \implies attacks(S, \alpha)).
\end{multline*}
\end{definition}


\begin{definition}
A set $S$ is a \textbf{complete extension} only if it is admissible and every acceptable argument with respect to $S$ belongs to it.
\begin{multline*}
    complete(S)\equiv_{def}adm(S)\wedge(\forall\alpha\in A\ acc(\alpha, S)\implies\\
    \implies (\alpha\in S)).
\end{multline*}
\end{definition}

\begin{definition}
A set $S$ is a \textbf{grounded extension} if it is the minimal (with respect to the set-theoretic inclusion) complete extension of $S$.
\begin{multline*}
    grounded(S)\equiv_{def}complete(S)\wedge(\forall T\subseteq A\ T\subset S\implies\\
    \implies\neg complete(T))
\end{multline*}
\end{definition}


There are inclusions between the defined extensions. Particularly, Dung proved that if an argumentation framework is well-founded, it has exactly one complete extension, which coincides with the grounded, preferred, and stable extensions.

\begin{definition}
An argumentation framework is \textbf{well-founded} only if there exists no infinite sequence $\alpha_0, \alpha_1,..., \alpha_n,...$ such that $\forall i$, $(\alpha_{i+1}, \alpha_i)\in R$.
\end{definition}

\begin{theorem}
\label{theo1}
Every well-founded argumentation framework has exactly one complete extension which is grounded, preferred and stable.
\end{theorem}

An argumentation framework with only one finite set of arguments is demonstrated in Example~\ref{ex1} below. It has only one preferred extension, which is also stable, complete and grounded. In contrast, the grounded extension is always unique and uncontroversial but may be empty if several admissible subsets are present, as shown in Example~\ref{ex2} adapted from \cite{caminada2007comparing}.

\begin{example}
\label{ex1}
Let $A=\{\alpha, \beta, \gamma\}$ and $(\alpha, \beta)\in R$. Then, if an argument $\gamma$ criticizes $\alpha$, i.e., $R=R\cup\{(\gamma, \alpha)\}$, the argument $\beta$ is returned to $A$ as defended. In this example, the arguments $\beta$ and $\gamma$ are acceptable, but not $\alpha$, since it is attacked without being defended. The acceptable arguments $\beta$ and $\gamma$ form several admissible argument sets: $\{\beta\}$, $\{\gamma\}$, and $\{\beta, \gamma\}$, but the preferred extension is the subset $\{\beta, \gamma\}$, which is also stable, complete and grounded.
\end{example}

\begin{example}
\label{ex2}
Continuing with Example~\ref{ex1}, let $R=R\cup\{(\delta, \alpha), (\gamma, \delta), (\delta, \gamma)\}$. This introduces a symmetric attack relation between $\gamma$ and $\delta$, creating a cycle of attacks. Thus, there are two conflict-free sets, $\{\beta, \gamma\}$ and $\{\beta, \delta\}$, both of which are preferred and stable, but there is no unique complete extension.
\end{example}


When multiple extensions exist, the concepts of credulous and skeptical acceptance of an argument are introduced \cite{boella2009dynamics}. However, we will later show that the formalization of peer review by argumentation frameworks yields well-founded extensions, making the distinction between credulous and skeptical semantics unnecessary in our case.

\subsection{Representation of Abstract Argumentation Frameworks in OWL DL}

We also outline the method for representing abstract argumentation frameworks in OWL DL presented in \cite{Baimuratov2023}. This method enables finding solutions to disputes using OWL reasoning engines. The process for generating these representations takes as input argumentation frameworks in JSON, and consists of three steps. The algorithm first generates basic OWL entities, including \texttt{owl:Classes} for the argument sets, \texttt{owl:NamedIndividuals} for the arguments, the \textit{attacks} relations as \texttt{owl:ObjectProperty}, and the inverse \textit{isAttackedBy} relation needed to generate admissible subsets. Next, the algorithm ``closes'' arguments with respect to the \textit{attacks} and \textit{isAttackedBy} relations to provide logical inference under the open world assumption. Finally, the algorithm generates definitions of conflict-free and admissible subsets of arguments for each input argument set.

\begin{figure*}[htb!]
    \centering
    \includegraphics[width=0.8\linewidth]{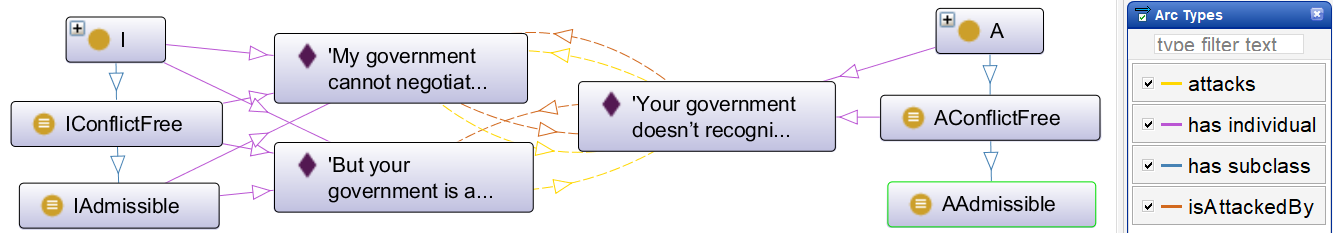}
    \caption{Example of a generated AAF representation in OWL}
    \label{fig:aaf_owl}
\end{figure*}

Once the abstract argumentation frameworks have been represented in OWL, their classification into admissible subsets of arguments can be automatically carried out using reasoning engines. This classification make it possible to identify which subsets of arguments correspond to preferred, stable, complete, or grounded extensions. As an example, \autoref{fig:aaf_owl} provides a visualization of the generated representation of Dung’s original AAF from \cite{Dung1995}, visualized using the OntoGraf tool\footnote{\url{https://protegewiki.stanford.edu/wiki/OntoGraf}}.

\section{Method}
\label{sec:method}

In this section, we introduce our method for facilitating dispute resolution in peer review using abstract argumentation frameworks. We begin by arguing that peer review can be viewed as an argumentative dispute, then proceed to formalize it using abstract argumentation frameworks. Next, we analyze the graph-theoretic properties of these frameworks and, finally, demonstrate how argumentation frameworks derived from peer reviews can be represented in OWL.

\subsection{Peer Review as an Argumentative Dispute}

We conceptualize peer review as an argumentative dispute between the authors of a manuscript submitted for publication and the reviewers evaluating it. By submitting their manuscript, the authors initiate an argumentative dispute in which they assert that their work meets all relevant requirements and deserves to be accepted. In other words, they express the claim $P$ --- the manuscript deserves acceptance, which serves as the root argument. By initiating this dispute, the authors expose their argument to three possible outcomes, depending on the reviewers' evaluation of the manuscript.

One possible outcome is that the reviewers agree with the authors’ assertion, resulting in the manuscript being accepted as is, indicating that there are no attacking arguments against the authors’ initial claim. Alternatively, the reviewers may argue that the manuscript fails to meet the required criteria for acceptance, leading to its rejection. In both scenarios, the dialogue between the authors and reviewers concludes after the first round of evaluation.

The third option on which we focus here involves the reviewers objecting to the authors’ claim by highlighting the need for revisions due to the manuscript's incomplete fulfillment of the acceptance criteria. We interpret the reviewers' objections, comments and recommendations as argumentative reasoning containing a negative point of view about the authors’ claim $P$, as well as criticizing the authors’ arguments. In this scenario, the dialog extends to at least one more round, where the authors respond to the reviewers with new arguments detailing the revisions made. The reviewers then reassess whether the manuscript meets the criteria after these corrections. We treat both the reviewers’ objections and the authors’ subsequent responses as attacking arguments within the dispute.

Thus, peer review can be seen as a single mixed dispute, where reviewers critique the authors' claim $P$ that the manuscript deserves acceptance and its supporting arguments, without presenting an opposing claim regarding $P$. In other words, the reviewers do not advocate for $\neg P$ (that the manuscript does not deserve acceptance) as a positive stance but focus solely on challenging the authors' claim that $P$ holds true.

\subsection{Formalizing Peer Review with Abstract Argumentation Frameworks}

Interpreting peer review as a single mixed dispute enables its formalization using abstract argumentation frameworks, leading to the creation of \textit{peer review argumentation frameworks}. In this approach, we do not evaluate individual peer review arguments based on their internal structure. Instead, the decision on whether the paper should be accepted as an outcome of peer review amounts to determining the acceptability of the root argument, which represents the overall evaluation of the manuscript.

Thus, we formalize peer review as an abstract argumentation framework and represent it using a JSON format based on the schema presented in \cite{Baimuratov2023}. In this schema, for each review party, there is a set of arguments, where every argument has a unique identifier and corresponding text. Attack relations between arguments are encoded as a list of pairs containing these argument identifiers. For the specific context of peer review, we extend this general schema by introducing two metadata properties of arguments: \textit{round} and \textit{number}. In the JSON format, these properties are incorporated within the argument identifiers. The identifiers follow the structure $<P.R.N>$, where $P$ denotes the review party, $R$ indicates the round number, and $N$ represents the argument number within the corresponding $<P.R>$ pair. By default, we include an empty argument for the authors’ party, representing the entire manuscript and serving as the root node of the argumentation framework. Listing~\ref{lst2} illustrates a JSON representation of the peer review argumentation framework using this extended schema. This framework consists of three argument sets corresponding to authors and two reviewers, along with four attack pairs.


\begin{lstlisting}[label=lst2, float=htb!, caption=Example of a peer review argumentation framework in JSON]
{
  "argument_sets": {
    "Reviewer_1": {
      "Reviewer_1.1.1": "However, being experts in their field the authors might not be aware that for readers less familiar with the metabolism/physiology of archaea, the examples are not always easy to follow..."
    },
    "Reviewer_2": {
      "Reviewer_2.1.1": "There is now available a great resource for the gene discovery... This should be discussed in the text..."
    },
    "Author": {
      "Author.2.1": "We have rephrased the four paragraphs where the referee found that the described examples are not always easy to follow.",
      "Author.2.2": "We have added a short outlook-type paragraph towards the end of the conclusions...",
      "Author.0.0": ""
    }
  },
  "attack_pairs": [
    [
      "Reviewer_1.1.1",
      "Author.0.0"
    ],
    [
      "Reviewer_2.1.1",
      "Author.0.0"
    ],
    [
      "Author.2.1",
      "Reviewer_1.1.1"
    ],
    [
      "Author.2.2",
      "Reviewer_2.1.1"
    ]
  ]
}
\end{lstlisting}

\subsection{Analysis of Peer Review Argumentation Frameworks}

The argumentation frameworks derived from the formalization of peer review possess distinct semantic and graph theoretic features due to the protocols of the peer review process. This subsection investigates these features and their implications for the computational complexity of determining argument acceptance within these frameworks.

In general, determining credulous acceptance of an argument with respect to a complete extension is $NP$-complete, while skeptical acceptance is $\Pi^p_2$-complete \cite{dvorak2018computational}. However, specific constraints on the attack relations in peer review argumentation frameworks reduce this computational complexity.

First, once a party in the peer review process presents an argument, that argument remains fixed within the current iteration and cannot retroactively attack new arguments introduced in subsequent iterations. Instead, it must be defended by arguments introduced in later iterations. Consequently, the attack relation in peer review argumentation frameworks is asymmetric:
\begin{equation*}
    \forall \alpha,\beta\in A\ (\alpha, \beta)\in R\implies (\beta, \alpha)\not\in R.
\end{equation*}

Next, except for the trivial case when there is only a single argument $\alpha_0$ representing the overall evaluation of the paper and no reviewers' arguments, the peer review argumentation frameworks cannot contain isolated nodes or disconnected subgraphs, since each new argument is introduced by attacking a previously existing one:
\begin{equation*}
    \forall \alpha_{i>0}\in A\ \exists \beta\in A\ (\alpha, \beta)\in R.
\end{equation*}

Moreover, parties' arguments can only attack arguments introduced in previous rounds. We denote by $\alpha^i$ an argument introduced in the $i$-th round, then:
\begin{equation*}
    \forall \alpha^i, \alpha^{i+1}\in A\ (\alpha^i, \alpha^{i+1})\not\in R.
\end{equation*}
Thus, peer review argumentation frameworks are acyclic. For directed acyclic graphs (DAGs), an extension can be found in linear time as $|A|+|R|$ and all decidability problems are $P$-hard.

Additionally, authors and reviewers produce only finite sets of arguments over a limited number of rounds. As a result, peer review argumentation frameworks are both finite and acyclic, which implies that they are well-founded.
\begin{theorem}
Review argumentation frameworks are well-founded.
\end{theorem}
According to Theorem~\ref{theo1}, they have a unique complete extension, which is also preferred, stable, and grounded. Therefore, resolving a peer review dispute reduces to identifying the unique complete extension and verifying whether the root node representing the overall evaluation of the paper is included within it.

Another notable feature of peer review argumentation frameworks is that they are $k$-partite graphs \cite{dunne2007computational}, where $k$ is the number of review parties.

\begin{definition}
An argumentation framework $<A, R>$ is \textbf{k-partite} if there exists a partition of $A$ into $k$ subsets $A_1,..., A_k$ such that
\begin{equation*}
    \forall(\alpha, \beta)\in R\ \beta\in A_i\implies \alpha\not\in A_i.
\end{equation*}   
\end{definition}

Review parties are not supposed to attack their own arguments, thus, the set of arguments $A$ is partitioned into disjoint subsets $A_1,..., A_k$ that represent each party's arguments, forming a $k$-partite directed graph. Despite this structure, for argumentation frameworks that are $k$-partite graphs with $k\geq 3$ the acceptance problem remains $NP$-complete. However, reducing the framework to a bipartite graph, for example, by partitioning the arguments into only two subsets: authots' and reviewers' arguments, results in $P$-time decidability.

Finally, peer review argumentation frameworks, being DAGs with a single starting point -- the node representing the overall evaluation of the paper -- essentially form trees. It is known from \cite{dunne2007computational}, that for argumentation frameworks with a tree-width bounded by a constant, both credulous and skeptical acceptance problems are decidable in linear time. Since the tree-width of a tree is 1, the peer review argumentation frameworks can again be resolved in linear time.

\subsection{Representing Peer Review Argumentation Frameworks in OWL DL}
\label{sec:representing}

Here, we demonstrate the application of the method for representing abstract argumentation frameworks in OWL DL from \cite{Baimuratov2023}, specifically tailored to the peer review process. In this approach, each review party is modeled as an \texttt{owl:Class}. Listing~\ref{lst3} shows the OWL classes for the review parties of the argumentation framework from Listing~\ref{lst2}, namely \textit{Author}, \textit{Reviewer\_1} and \textit{Reviewer\_2}, in Manchester syntax.

\begin{lstlisting}[label=lst3, float=htb!, caption=Review parties in OWL]
Class: <onto#Author>

  SubClassOf:
    owl:Thing

Class: <onto#Reviewer_1>

  SubClassOf:
    owl:Thing

Class: <onto#Reviewer_2>

  SubClassOf:
    owl:Thing
\end{lstlisting}

Each argument of the review parties is represented as \texttt{owl:Named\-Individual}, with the argument text captured using a custom \texttt{owl:Annotation\-Property} named \textit{text}. The association of each argument with its respective review party is indicated by the \texttt{rdf:type} relation. Attack relations between arguments are asserted using the \texttt{owl:ObjectProperty} \textit{attacks} and its inverse \textit{isAttackedBy}. Additionally, specific properties introduced for the peer review scenario, \textit{round} and \textit{number}, are also represented as  \texttt{owl:AnnotationProperty}. To ensure logical inference under the open world assumption, each individual is ``closed'' with respect to the \textit{attacks} and \textit{isAttackedBy} relations. Listing~\ref{lst4} illustrates how a peer review argument from Listing~\ref{lst2} is represented in OWL.

\begin{lstlisting}[label=lst4, float=htb!, caption = Representation of a peer review argument in OWL]
Individual: <onto#reviewer_11>

  Annotations:
    <onto#number> "1"^^xsd:string,
    <onto#round> "1"^^xsd:string,
    <onto#text> "However, being experts in their field the authors might not be aware that for readers less familiar with the metabolism/physiology of archaea, the examples are not always easy to follow..."^^xsd:string

  Types:
    <onto#Reviewer_1>
    <onto#attacks> only({onto#author_3>}),
    < onto#isAttackedBy> only({<onto#author_1>})

  Facts:  
    <onto#attacks><onto#author_3>
    <onto#isAttackedBy> <onto#author_1>
\end{lstlisting}

Then, we generate conflict-free and admissible argument subsets for each review party. Listing~\ref{lst5} presents the declarations of these subsets for the \textit{Author} party from Listing~\ref{lst2}.

\begin{lstlisting}[label=lst5, float=htb!, caption = Conflict-free and admissible subsets of authors' arguments in OWL]
Class: <onto#AuthorConflictFree>

  EquivalentTo:
    <onto#Author>
    and (<onto#attacks> only (<onto#Reviewer_1> or <onto#Reviewer_2>))

Class: <onto#AuthorAdmissible>

  EquivalentTo:
    <onto#AuthorConflictFree>
    and (<onto#isAttackedBy> only(<onto#isAttackedBy> some <onto#AuthorConflictFree >) )
\end{lstlisting}

By formalizing peer review in the OWL DL language, OWL reasoners such as Pellet \cite{Sirin2007} can be employed to determine the acceptability of each party’s arguments. The reasoner classifies all arguments from a given review party $A_i$, assigning them to the corresponding subset and thereby constructing the complete extension. Notably, reasoning in Description Logics is known to be PSpace-complete\footnote{\url{http://www.cs.man.ac.uk/~ezolin/dl/}}, making it computationally feasible for the scale of argumentation frameworks typically encountered in peer review.

Therefore, the proposed workflow for resolving disputes in peer review involves four key steps: 1) formalizing a peer review as an abstract argumentation framework in JSON format, 2) generating an OWL DL representation of the formalized peer review, 3) applying a reasoning engine to determine the acceptability of the arguments, and 4) verifying whether the argument representing the overall evaluation of the paper is acceptable.
\section{Evaluation}
\label{sec:eval}

To evaluate our method for resolving disputes in peer review using abstract argumentation frameworks, we manually formalized a corpus of scientific peer reviews and applied the proposed approach to resolve each of them. Additionally, we computed and analyzed statistical properties of the resulting peer review argumentation frameworks.

\paragraph{Data}

We used the open peer review corpus provided by \cite{Miłkowski2022} to annotate abstract argumentation frameworks. This corpus includes 123 peer-reviewed articles published in MDPI journals, accessible as of June 16, 2022. The reviews are presented in their original formats, as uploaded by reviewers through the MDPI editorial system, primarily in PDF, TXT, and DOCX formats. Additionally, the corpus contains metadata on specific reviews, author responses, and article information, all available in JSON format. The corpus is licensed under the Creative Commons Attribution 4.0 (CC BY) license.

\paragraph{Annotation}

To evaluate the proposed method for dispute resolution in scientific peer review, we manually annotated the selected corpus of peer reviews with abstract argumentation frameworks. This involved identifying individual arguments and the attack relations between them within each review, and storing the resulting frameworks in the JSON format described in Subsection~\ref{sec:representing}. Some reviews, however, were incomplete, preventing full reconstruction of argumentation chains and thus could not be formalized. In total, we successfully compiled 88 abstract argumentation frameworks from the corpus.

\paragraph{Implementation}

\begin{figure*}[htb!]
    \centering
    \includegraphics[width=\textwidth]{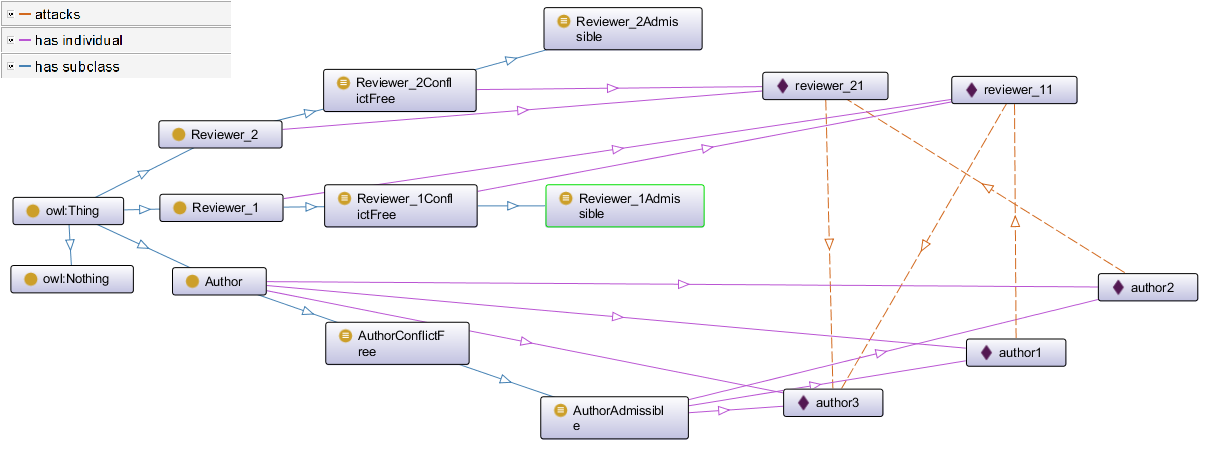}
    \caption{Example of a generated peer review argumentation framework in OWL}
    \label{fig1}
\end{figure*}

A prototype was developed to generate OWL representations from the annotated peer reviews using Python and the Owlready2 library \cite{Owlready}. The Pellet reasoner \cite{Sirin2007} was employed to classify arguments into admissible subsets. The prototype successfully generated OWL representations for all annotated peer reviews and correctly identified a complete extension for each, thereby validating the proposed method. Listing~\ref{lst7} shows the reasoning logs from running Pellet on the OWL representation of the peer review argumentation framework introduced in Listing~\ref{lst2}. Figure~\ref{fig1} presents a visualization of this framework after classification. The experiments were conducted on a machine running Windows 10, with an Intel Core i5-1035G1 CPU and 8GB of RAM. The developed software, along with the annotated peer reviews and their corresponding OWL representations, is available in the GitHub repository
\footnote{\url{https://anonymous.4open.science/r/ReviewArgumentationFramework-E523}}

\begin{lstlisting}[label=lst7, float=htb!, caption = Classification of review arguments by Pellet reasoner]
* Owlready2 * Pellet took 1.5788683891296387 seconds
* Owlready2 * Pellet output:

 http://www.w3.org/2002/07/owl#Thing
    onto#Author
       onto#AuthorConflictFree
          onto#AuthorAdmissible - (onto#author3, onto#author2, onto#author1)
    onto#Reviewer_1
       onto#Reviewer_1ConflictFree - (onto#reviewer_11)
          onto#Reviewer_1Admissible
    onto#Reviewer_2
       onto#Reviewer_2ConflictFree - (onto#reviewer_21)
          onto#Reviewer_2Admissible
\end{lstlisting}

\paragraph{Results}

After identifying the complete extensions, we examined each peer review argumentation framework to determine whether the root argument representing the overall evaluation of the authors’ manuscript was acceptable. As a result, only 53.4\% of the papers were found to be acceptable. In addition, we computed various statistical features of the generated OWL representations. Specifically, we analyzed the following characteristics of each argumentation framework: the number of review parties; the number of arguments presented by the authors; the total and average (per reviewer) number of arguments made by reviewers; the average length of attack chains between arguments; the number of acceptable arguments from the authors; the total and average number of acceptable arguments from reviewers; and the average time required to convert the peer review framework from JSON to OWL and to infer its complete extension. A summary of these characteristics is presented in Table~\ref{tab2}.

\begin{table}[htb!]
\centering
\begin{tabular}{p{5.5cm}c}
\toprule
Feature & Value \\
\midrule
Papers acceptability rate & 53.4\% \\
Number of review parties & $3.41\pm 0.71$ \\
Number of authors' arguments & $22.24\pm 14.62$ \\
Total number of reviewers' arguments & $21.99\pm 14.66$ \\
Average number of reviewers' arguments & $9.35\pm 6.24$ \\
Length of attack chains & $2.95\pm 0.08$ \\
Number of authors' acceptable arguments & $21.77\pm 14.59$ \\
Total number of reviewers' acceptable arguments & $1.06\pm 2.2$ \\
Average number of reviewers' acceptable arguments & $0.42\pm 0.66$ \\
Time (s) & $0.002\pm 0.005$ \\
\bottomrule
\end{tabular}
\caption{Features of resolved peer review argumentation frameworks}
\label{tab2}
\end{table}

\paragraph{Interpretation}

As anticipated, the number of authors' arguments, the total number of reviewers' arguments, and the number of authors' acceptable arguments are closely aligned, as authors typically respond to each argument raised by the reviewers. Given that the total number of reviewers' acceptable arguments is more than one and that a single acceptable reviewer argument is sufficient to make the paper unacceptable, the papers acceptability rate of 53.4\% (47 out of the 88 annotated peer reviews) is consistent with expectations. Considering that all analyzed papers were ultimately accepted by MDPI, this finding suggests that in 46.6\% of cases, editors accepted papers without fully addressing all reviewers' concerns. While some of these arguments might have been minor, integrating our approach could enhance the quality of published work. Additionally, the average length of attack chains consistently is around three, suggesting that the most common attack pattern is \textit{Paper $\leftarrow$ Reviewer's Argument $\leftarrow$ Authors' Argument}. Other metrics, such as the total number of arguments from both authors and reviewers, the number of accepted arguments, and the overall paper acceptance rate, display considerable variability.
\section{Conclusion}
\label{sec:conc}

In this study, we conceptualized peer review as a single mixed argumentative dispute between the authors of a submitted manuscript and the reviewers evaluating it. We demonstrated the applicability of abstract argumentation frameworks to the peer review process and analyzed the resulting peer review argumentation frameworks from semantic, graph-theoretic, and computational perspectives. In particular, we argued that these frameworks are well-founded and are decidable in linear time. Building on this, we adapted an existing method for generating OWL DL representations of argumentation frameworks to the specific context of peer review. This adaptation enables the use of OWL reasoning engines to determine peer review argument acceptability. Thus, our proposed workflow for resolving disputes in peer review involves: 1) annotating a peer review with abstract argumentation frameworks, 2) generating an OWL representation of the annotated framework, 3) classifying arguments using OWL reasoning, and 4) determining whether the argument representing the manuscript’s overall evaluation is acceptable. To validate the method, we annotated a corpus of peer reviews with abstract argumentation frameworks, implemented a prototype to generate their OWL representations, and identified their complete extensions via reasoning. We also analyzed features of the resulting frameworks. Our findings suggest that integrating this approach could significantly enhance the peer review process by providing a more transparent and rigorous way to account for reviewers’ arguments.

\paragraph{Limitations and Future Work}

Currently, our approach relies on the manual annotation of peer reviews, a time-consuming process that is susceptible to human error. In future work, we plan to incorporate argumentation mining techniques by leveraging machine learning models to automate the extraction of arguments and their attack relations. Alternatively, dedicated user interfaces could be developed for peer review platforms to guide reviewers in structuring their feedback according to abstract argumentation frameworks. Such interfaces could also highlight acceptable arguments, thereby supporting editors and metareviewers in making more transparent and informed decisions. 

\bibliographystyle{kr}
\bibliography{kr-sample}

\end{document}